\documentclass[conference]{IEEEtran}
\IEEEoverridecommandlockouts
\usepackage{cite}
\usepackage{amsmath,amssymb,amsfonts}
\usepackage{algorithmic,comment}
\usepackage{graphicx}
\usepackage{textcomp}
\usepackage{xcolor}
\usepackage{booktabs}
\usepackage{multirow} 

\usepackage[hidelinks,colorlinks=true,linkcolor=blue,citecolor=cyan]{hyperref}

\usepackage{enumitem}

\usepackage{balance}

\usepackage{setspace}
\def\BibTeX{{\rm B\kern-.05em{\sc i\kern-.025em b}\kern-.08em
    T\kern-.1667em\lower.7ex\hbox{E}\kern-.125emX}}
\begin{document}

\title{BladderFormer: A Streaming Transformer for Real-Time Urological State Monitoring
 \vspace{-.3em}
}

\author{
Chengwei Zhou \quad
Steve Majerus \quad
Gourav Datta \\
Case Western Reserve University, Cleveland, USA  \\
{\tt\small \{chengwei.zhou, steve.majerus, gourav.datta\}@case.edu}
\vspace{-4mm}
}

\maketitle

\begin{abstract}
Bladder pressure monitoring systems are increasingly vital in diagnosing and managing urinary tract dysfunction. Existing solutions rely heavily on hand-crafted features and shallow classifiers, limiting their adaptability to complex signal dynamics. We propose a one-layer streaming transformer model for real-time classification of bladder pressure states, operating on wavelet-transformed representations of raw time-series data. Our model incorporates temporal multi-head self-attention and state caching, enabling efficient online inference with high adaptability. Trained on a dataset of 91 patients with 20,000-80,000 samples each, our method demonstrates improved accuracy, higher energy- and latency-efficiency. Implementation considerations for edge deployment on low-power hardware, such as edge graphical processing units (GPU) and micro-controllers, are also discussed.
\end{abstract}


\vspace{-2mm}

\section{Introduction}

Continuous monitoring and classification of bladder pressure dynamics is a critical enabler for intelligent, closed-loop control in urinary management systems~\cite{majerus2021feasibility,park2018detecting}. Accurate real-time identification of physiological states such as detrusor overactivity, voluntary voiding, or abdominal interference plays a central role in improving patient outcomes, reducing catheter dependency, and advancing neuroprosthetic solutions~\cite{abelson2019ambulatory,soebadi2019wireless}. These diagnostic decisions rely on precise biosignal interpretation under tight latency and energy constraints—requirements unmet by conventional machine learning models that lack support for on-board, real-time inference~\cite{karam2016real,karam2017tunable}. Traditional urodynamic classification systems have predominantly relied on hand-engineered features and shallow classifiers such as logistic regression or artificial neural network (ANN) classifiers~\cite{hobbs2022machine,wang2021pattern,melgaard2014detecting}. Although computationally efficient, these approaches exhibit limited robustness across patient populations and perform poorly when generalized to new signal variations~\cite{frenkl2011variability,radley2001conventional}. Moreover, they fail to extract rich temporal features from time-varying bladder dynamics and are restricted to single-label classification, limiting their ability to differentiate among multiple physiological bladder states~\cite{majerus2024realtime,hobbs2022machine}.

Recent advancements in attention-based deep learning, particularly transformer models, have revolutionized sequential modeling in natural language processing and vision~\cite{vaswani2017attention,dosovitskiy2020image}. Their application to biomedical time-series analysis—such as ECG arrhythmia detection~\cite{natarajan2020wide,liu2021inter}, EEG seizure classification~\cite{song2021attend,kostas2021bendr}, and respiration pattern analysis~\cite{ahmed2019heartnet,wang2020transformers}—has shown superior performance in capturing long-range dependencies. Nonetheless, these transformer-based models typically involve multiple stacked layers and operate on long input sequences in batch mode, making them impractical for streaming inference on constrained hardware platforms such as wearable devices or microcontrollers~\cite{strubell2019energy,qin2022efficient,wang2021linformer}.

To bridge this gap, we propose a novel streaming transformer architecture tailored for ultra-low-latency bladder pressure classification. Our approach divides incoming physiological signals into short, fixed-length segments (e.g., 8 samples) and performs self-attention within each segment for fine-grained local integration. To capture longer-term temporal dependencies without introducing recurrence, we augment this with a lightweight context vector caching mechanism that stores pooled embeddings from previous segments. This enables causal attention across time while bounding computation and memory overhead. Unlike recurrent neural networks (RNNs) or long short-term memory (LSTM) models, which rely on persistent hidden states and sequential updates, our model performs parallel attention over current segments and cached past context, avoiding the computational bottlenecks of recurrence. Despite this segment-based processing, real-time performance is retained, as each segment spans only 0.8 seconds of data at a 10~Hz sampling rate. Moreover, the classification task is grounded in clinically validated annotations from expert urologists, where each time sample may belong to multiple overlapping physiological categories: \textit{None}, \textit{Overactivity}, \textit{Voiding}, and \textit{Abdominal} Activity. This multi-label setting reflects the complex, overlapping nature of bladder dynamics, and is not addressed by prior approaches that simplify the task to binary classification. Our method natively supports multi-label inference while preserving temporal resolution and minimizing energy consumption. The resulting architecture—comprising a single-layer transformer with segment-wise and cross-segment attention—achieves 83.25\% classification accuracy, outperforming previous ANN-based methods while remaining deployable on low-power microcontrollers. The low parameter count, shallow depth, and fixed-cost attention ensure real-time inference suitable for wearable or implantable autonomous bladder management systems.

\section{Related Work}

Early approaches to automated bladder pressure analysis emphasized frequency-domain features and conventional signal processing techniques. Cullingsworth et al.\cite{cullingsworth2018automated} and Niederhauser et al.\cite{niederhauser2018detection} demonstrated that spectral analysis and rhythmic contraction profiling could help stratify patients with lower urinary tract dysfunction, laying important groundwork for automated characterization of detrusor overactivity. Expanding on this direction, Karam et al.~\cite{karam2016real} introduced the Context-Aware Thresholding (CAT) algorithm, one of the first published systems for automated urological event detection from single-channel vesical pressure. Their approach combined wavelet-based adaptive thresholding with efficient signal processing to detect bladder voiding contractions with 97\% accuracy within 1 second of onset. However, the model supported only binary detection and did not differentiate among multiple event types.

The introduction of machine learning methods led to more nuanced classification strategies. Wang et al.\cite{wang2021pattern} employed manifold learning with dynamic time warping to compare 6-second detrusor pressure (PDET) segments against a library of canonical overactivity patterns, achieving 81.4\% accuracy in identifying detrusor overactivity. Hobbs et al.\cite{hobbs2022machine} extended this by evaluating both time-domain and frequency-domain features with support vector machine classifiers. Their results showed that single-channel time-domain features could yield 85.4\% sensitivity, while combining three-channel frequency features led to 92.9\% specificity. Despite their promise, these methods were limited by their reliance on fixed-length (e.g., 60-second) input windows and their focus on specific event classes, restricting their applicability in low-latency or multi-event settings. Other classical methods using support vector machines and decision trees have also been proposed~\cite{melgaard2014detecting}, but these typically depend on pre-defined features and static, segment-level classification, lacking temporal modeling capabilities. More recently, Majerus et al.~\cite{majerus2024realtime} presented a machine learning pipeline combining lifting wavelet transforms with 16 scalar features processed by a compact 4-layer ANN. Their system achieved 79\% accuracy on real-time classification tasks deployed on microcontrollers operating at 10 Hz, illustrating feasibility for embedded applications. However, the model operated on isolated segments, without temporal adaptation across samples, and was constrained to single-label predictions per segment, limiting its utility for overlapping or evolving physiological states. In contrast, our streaming transformer integrates short-range temporal dependencies via self-attention, capturing bladder activity transitions more effectively. 


\section{Proposed Method}

\subsection{Segment-Level Attention for Local Temporal Modeling}

Our model operates on fixed-length segments of eight consecutive bladder pressure samples, capturing local temporal dependencies while maintaining bounded computational cost. Unlike recurrent models that require sequential processing, our transformer architecture processes all samples within each segment in parallel, introducing only a 0.8-second buffering delay at 10~Hz sampling. This 8-sample window balances temporal resolution with computational efficiency, capturing typical bladder pressure transients (0.5-1.0 seconds) while enabling real-time urological monitoring.

\begin{figure}
    \centering
    \includegraphics[trim={230 293 180 190}, clip, width=\linewidth]{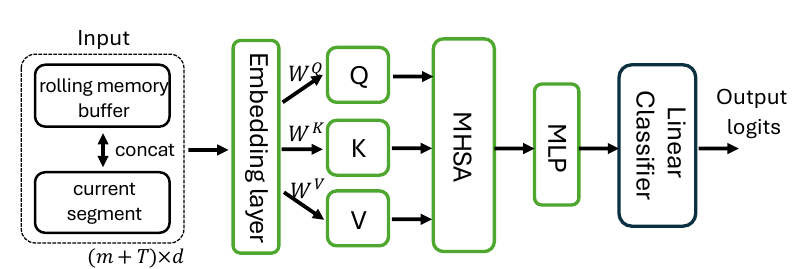}
    \caption{Architecture of the cross-segment caching mechanism for long-term temporal adaptation with single-layer streaming transformer, incorporating with Multi-Head Self-Attention (MHSA) and multi-layer perceptron (MLP) layers, with key components (Q, K, V) for attention computation.}
    \vspace{-4mm}
    \label{fig:archi}
\end{figure}

\noindent
\textbf{Feature Extraction:} Let $\mathbf{x}{=}[x_1, x_2, \dots, x_{512}]{\in}\mathbb{R}^{512}$ denote the latest 512 samples stored in a \textit{first in first out} (FIFO) buffer of scalar bladder pressure values. Each buffer is decomposed using a 5-level Lifting Wavelet Transform with Daubechies-4 wavelets: $\text{LWT}_5(\mathbf{x}) = \{A_5, A_4, A_3, A_2, A_1, D_5, D_4, D_3, D_2, D_1\}$, where $A_i \in \mathbb{R}^{\frac{512}{2^{j}}}$ are the approximation coefficients at level $i$ for $i = 1, 2, 3, 4, 5$, and $D_j \in \mathbb{R}^{\frac{512}{2^{j}}}$ are the detail coefficients at level $j$ for $j = 1, 2, 3, 4, 5$. At each time step, only the most recent coefficient from each sub-band is retained:
\begin{align}
    \mathbf{c} = [& A_5[-1], A_4[-1], A_3[-1], A_2[-1], A_1[-1], \notag \\
                 & D_5[-1], D_4[-1], D_3[-1], D_2[-1], D_1[-1]] \in \mathbb{R}^{10} \notag
\end{align}
To enrich the feature representation, ratios between retained wavelet coefficients are computed as $\mathbf{r} = \left[\frac{c_i}{c_i}\right]_{i \in \mathcal{I}} \in \mathbb{R}^5$, where $\mathcal{I}$ represents 5 selected coefficient pairs chosen correspondingly
(from the 45 possible pairwise combinations)
to capture physiologically relevant frequency relationships that distinguish between bladder contractions, abdominal artifacts, and baseline activity patterns.
The feature vector for sample $t$ combines the original sample value, retained coefficients, and their ratios: $\mathbf{e}_t = [x_t, \mathbf{c}, \mathbf{r}] \in \mathbb{R}^{16}$.

\noindent
\textbf{Segment-wise Attention:} For each new sample, we group $T{=}8$ consecutive feature vectors to form segment matrices $\mathbf{E}^{(t)} = [\mathbf{e}_{t-7}, \mathbf{e}_{t-6}, \dots, \mathbf{e}_t] \in \mathbb{R}^{8 \times 16}$ for transformer processing. We first apply a linear projection to map the 16-dimensional features to a 64-dimensional representation:
\begin{equation}
    \mathbf{E}_{\text{proj}}^{(t)} = \mathbf{E}^{(t)} \mathbf{W}_{\text{proj}} \in \mathbb{R}^{8 \times 64}
\end{equation}
We then apply standard Multi-Head Self-Attention (MHSA) over each 8-sample segment:
\begin{align}
    \mathbf{Q}^{(t)}&{=}\mathbf{E}_{\text{proj}}^{(t)} \mathbf{W}_Q, \
    \mathbf{K}^{(t)}{=}\mathbf{E}_{\text{proj}}^{(t)} \mathbf{W}_K, \
    \mathbf{V}^{(t)}{=}\mathbf{E}_{\text{proj}}^{(t)} \mathbf{W}_V{\in}\mathbb{R}^{8{\times}64}, \notag \\
    \mathbf{Z}^{(t)} &{=}\text{softmax}\left( \frac{\mathbf{Q}^{(t)} \mathbf{K}^{(t)\top}}{\sqrt{64}} \right) \mathbf{V}^{(t)} \in \mathbb{R}^{8{\times}64}.
\end{align}
The resulting self-attended representation is passed through a position-wise multi-layer perceptron (MLP) network, followed by mean pooling across the temporal dimension:
\begin{align}
    \mathbf{H}^{(t)} &= \text{ReLU}(\mathbf{Z}^{(t)} \mathbf{W}_1 + \mathbf{b}_1) \in \mathbb{R}^{8{\times}64}, \\
    \mathbf{O}^{(t)} &= \mathbf{H}^{(t)} \mathbf{W}_2 + \mathbf{b}_2 \in \mathbb{R}^{8{\times}16}, \\
    \hat{y}^{(t)} &= \text{softmax}\left(W_o \cdot \frac{1}{8}\sum_{i=1}^{8} \mathbf{O}^{(t)}_i + \mathbf{b}_o\right),
\end{align}
predicting one of the four bladder states.
For multi-label bladder state prediction, we replace the softmax activation with an element-wise sigmoid function to enable independent probability estimation for each class, allowing simultaneous detection of overlapping physiological states (e.g., concurrent abdominal activity during voiding). This localized attention mechanism is computationally efficient and allows the model to learn temporal structure across samples within each short segment. However, it alone is insufficient to model long-term dependencies present in physiological dynamics, which motivates our cross-segment state caching strategy as described below. Our model architecture is illustrated in Fig. \ref{fig:archi}.

\subsection{Cross-Segment Caching for Temporal Adaptation}

To capture slower-evolving physiological changes across time, we introduce a causal memory mechanism that integrates context from previous segments without incurring the overhead of full recurrent processing. Instead of retaining all past token embeddings, we store a lightweight representation for each segment, enabling efficient global adaptation. For each segment $t$, we compute a mean-pooled context vector:
\begin{equation}
    \mathbf{r}^{(t)} = \frac{1}{8}\sum_{i=1}^{8} \mathbf{E}^{(t)}_i \in \mathbb{R}^{1 \times 16}.
\end{equation}
We maintain a rolling memory buffer of the last $m$ segment embeddings, where $m{=}8$ in our experiments:
\begin{equation}
    \mathbf{R}^{(t-1)} = [\mathbf{r}^{(t-m)}, \ldots, \mathbf{r}^{(t-1)}] \in \mathbb{R}^{m \times 16}.
\end{equation}
At the current segment $t$, we concatenate this memory with $\mathbf{E}^{(t)}$ and then apply a linear projection and self-attention over the combined set:
\begin{align}
    \tilde{\mathbf{E}}^{(t)}&{=}[\mathbf{R}^{(t-1)}; \mathbf{E}^{(t)}]{\in}\mathbb{R}^{16{\times}16}, \ 
    \tilde{\mathbf{E}}_{\text{proj}}^{(t)}{=}\tilde{\mathbf{E}}^{(t)} \mathbf{W}_{\text{proj}}{\in}\mathbb{R}^{16{\times}64}, \notag \\
    \tilde{\mathbf{Q}}&{=} \tilde{\mathbf{E}}_{\text{proj}}^{(t)} \mathbf{W}_Q, \
    \tilde{\mathbf{K}}{=} \tilde{\mathbf{E}}_{\text{proj}}^{(t)} \mathbf{W}_K, \
    \tilde{\mathbf{V}}{=} \tilde{\mathbf{E}}_{\text{proj}}^{(t)} \mathbf{W}_V{\in}\mathbb{R}^{16{\times}64}, \notag \\
    \tilde{\mathbf{Z}}^{(t)}&{=} \text{softmax}\left( \frac{\tilde{\mathbf{Q}} \tilde{\mathbf{K}}^\top}{\sqrt{64}} \right) \tilde{\mathbf{V}} \in \mathbb{R}^{16{\times}64}.
\end{align}
$\tilde{\mathbf{Z}}^{(t)}$ is then forwarded to the succeeding MLP and classifier layers (similar to Eqs. 3-5), while the memory bank $\mathbf{R}^{(t)}$ is updated by appending $\mathbf{r}^{(t)}$ and discarding the oldest entry. 
With memory buffer size $m{=}8$, this retains approximately 6.4 seconds of historical context, which may be sufficient to model the temporal dependencies observed in bladder filling dynamics~\cite{niederhauser2018detection} while maintaining efficient memory usage. 



\begin{table}[!t]
\centering
\small
\caption{Compute, memory, and estimated latency per segment ($d$=64, $n$=8) on \textsc{NVIDIA Jetson P3450}. L denotes latency.}
\label{tab:flops_latency}
\vspace{-2mm}
\begin{tabular}{p{2.6cm}ccc}
\toprule
\textbf{Component} & \textbf{FLOPs (M)} & \textbf{Mem. (kB)} & \textbf{L (ms)} \\
\midrule
Embedding layer & 0.016 & 1 & 1.71 \\
Q/K/V projections & 0.196 & 12 & 2.91 \\
MHSA & 0.033 & 0 & 10.84 \\
Feedforward MLP & 0.524 & 32 & 4.65 \\
Classifier head & 0.0004 & 0.25 & 0.90 \\
\midrule
\textbf{Total} & \textbf{0.77} & \textbf{45.25} & \textbf{21.01} \\
\bottomrule
\end{tabular}
\vspace{-3mm}
\end{table}

\subsection{Resource Efficiency and Embedded Deployment}

With a single transformer encoder layer processing fixed input windows of $(m+T){=}16$ tokens, our architecture achieves high efficiency in both compute and memory usage. Table~\ref{tab:flops_latency} illustrates the estimated floating-point operations (FLOPs), memory requirements, and latency for each module, assuming an embedding dimension of $d{=}16$. Using 8-bit quantized weights and fixed-point activations, the total model footprint remains ${\sim}$45.2~kB, comfortably fitting within the 256-512~kB RAM and 1~MB Flash constraints of ARM Cortex-M4/M7 chips. The memory buffer adds minimal overhead (0.125~kB for $m=8, d=16$). This design provides three key advantages for real-time deployment. First, it approximates recurrent temporal memory through pooled cross-segment context, eliminating the computational overhead of sequential processing. Second, it enforces causal flow by preventing access to future segments, ensuring streaming compatibility. Third, by maintaining a fixed-size memory buffer, computational cost remains bounded per segment regardless of total sequence length processed. These properties make our architecture particularly amenable to hardware acceleration (via dedicated attention units or tensor processing cores)  and well-suited for edge devices with strict latency requirements.

Inference per segment executes in 21~ms on an NVIDIA Jetson module (see Table \ref{tab:flops_latency}) using unoptimized PyTorch, supporting real-time 10~Hz throughput. The ML computation adds only 33 ${\mu}$W of power overhead—negligible compared to the Jetson's 5-10W baseline consumption. This minimal footprint demonstrates suitability for ARM Cortex microcontrollers, where total system power would be 50-100mW (13.5-27mA at 3.7V). With duty-cycled operation, average consumption reduces to 1.4-2.7mA, enabling 7-10 days of continuous operation on a 500mAh battery. Since ML inference represents ${<}$0.7\% of the total power budget on the energy-intensive Jetson platform, computational overhead remains negligible on purpose-built low-power systems, making real-time bladder monitoring practical for ambulatory patient care.


\section{Dataset and Preprocessing}
\begin{figure}
    \centering
    \includegraphics[trim={150 200 190 170}, clip, width=1\linewidth]{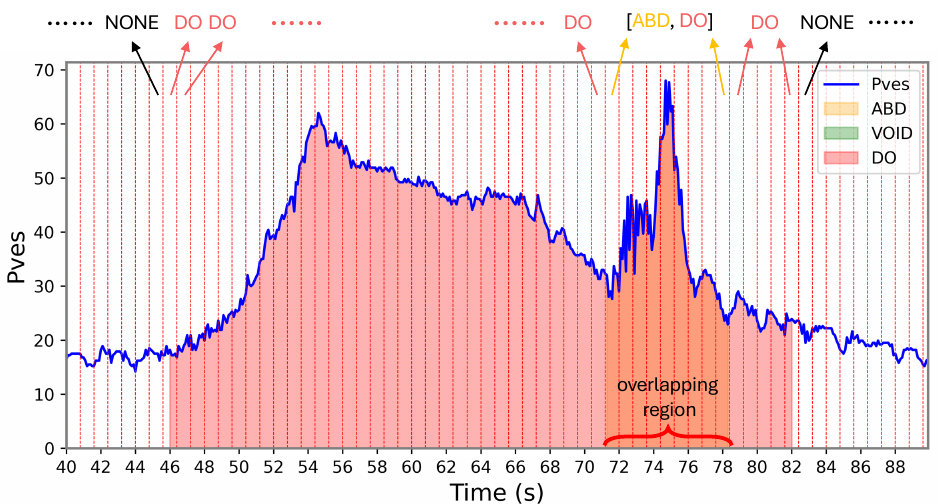}
    \caption{Example of segment-based labeling for a single interpolated bladder pressure (Pves) trace. Vertical dotted lines represent non-overlapping 0.8-second segments (8 samples at 10 Hz), each assigned a class label based on the event(s) within the window. The region enclosed by the \textcolor{red}{red brace (approximately 71s to 78s)} illustrates an overlapping zone where both DO and ABD events are present. In the multilabel classification setting, these segments are labeled with both classes simultaneously (i.e., [ABD, DO]). }
\label{fig:segmentation}
\vspace{-3mm}
\end{figure}

\begin{table*}[ht]
\centering
\scriptsize
\caption{Performance comparison between a baseline 4-layer ANN, and our Segment-Level Transformer, and Streaming Transformer models. $N$ denotes the number of 0.8-second segments from each class.}
\vspace{-2mm}
\label{table:performance}
\begin{tabular}{ll|rrrrrr|rrrrrr|rrrrrr}
\toprule
\multirow{4}{*}{\textbf{Class}} & \multirow{4}{*}{\textbf{$N$}} 
& \multicolumn{6}{c|}{\textbf{Baseline}} 
& \multicolumn{6}{c|}{\textbf{Segment-Level Transformer}} 
& \multicolumn{6}{c}{\textbf{Streaming Transformer}} \\
\cmidrule(lr){3-8} \cmidrule(lr){9-14} \cmidrule(lr){15-20}
& & \textbf{Acc.} & \textbf{F1} & \textbf{TP} & \textbf{TN} & \textbf{FP} & \textbf{FN}
  & \textbf{Acc.} & \textbf{F1} & \textbf{TP} & \textbf{TN} & \textbf{FP} & \textbf{FN}
  & \textbf{Acc.} & \textbf{F1} & \textbf{TP} & \textbf{TN} & \textbf{FP} & \textbf{FN} \\
\midrule
\multicolumn{7}{l}{\textit{single-label:}} \\
ABD   & 720   & 93.1 & 54.8 & 64.4 & 95.1 & 4.9 & 35.6   & 90.4 & 60.4  & 58.2 & 97.3 & 2.7 & 41.8  & 95.2 & 54.1 & 62.1 & 98.0 & 2.0 & 37.9 \\
DO    & 554   & 91.9 & 6.7 & 60.8 & 95.5 & 4.9 &  33.7    & 90.2 & 21.4 & 60.7 & 94.8 &5.2 &39.3  & 92.7 & 15.8 & 58.5 & 95.2 & 4.8 & 41.5 \\
VOID  & 1,240 & 85.2 & 62.4 & 72.3 & 87.8 & 12.2 & 27.7  & 87.2 & 63.5  & 75.4 & 87.9 & 12.1 & 24.6  & 89.0 & 67.5 & 76.3 & 90.2 & 9.8 & 23.7 \\
\textbf{Overall} & \textbf{12,440} 
& \textbf{78.8} & - & - & - & - & - 
& \textbf{79.2} & - & - & - & - & - 
& \textbf{82.12} & - & - & - & - & - \\
\midrule
\multicolumn{7}{l}{\textit{multi-label:}} \\
ABD   & 720   & 92.1 & 51.3  & 56.2 & 96.5 & 3.5 & 43.8  & 93.6 & 49.9 & 56.2 & 96.8 & 3.2 & 43.8 & 94.5 & 59.1 & 47.6 & 96.7 & 3.3 & 52.4   \\
DO    & 567   & 89.1 & 6.1 & 49.7 & 98.1 & 1.9 & 50.3  &  90.4 & 12.1 & 51.3 & 97.4 & 2.6 & 48.7 & 91.8 & 12.5 & 53.8 & 97.2 & 2.8 & 46.2  \\
VOID  & 1243 & 87.1 & 48.7 & 61.8 & 89.9 & 10.1 & 38.2  & 91.0 & 38.0 & 56.4 & 95.2 & 4.8 & 43.6 & 92.1 & 41.0 & 58.8 & 94.8 & 5.2 & 41.2   \\
\textbf{Overall} & \textbf{12,456} 
& \textbf{82.56} & - & - & - & - & - 
& \textbf{84.34} & - & - & - & - & - 
& \textbf{86.9} & - & - & - & - & - \\
\bottomrule
\end{tabular}
\vspace{-3mm}
\end{table*}

We evaluate our models on the same clinical dataset used by a recent baseline method~\cite{majerus2024realtime}, consisting of 108 bladder filling traces collected from human subjects at the Cleveland Clinic. From this dataset, we curate 91 distinct patient sessions for training, each containing approximately 20,000 to 80,000 samples, with annotations indicating one or more of four bladder activity states, as illustrated in Fig.~\ref{fig:segmentation}. 
For training, we construct input segments using a sliding window of eight consecutive wavelet-transformed samples with stride one, maximizing label utilization. For evaluation, we use non-overlapping segmentation following the baseline method \cite{majerus2024realtime}. The features are normalized before sending to the model. 

Expert urologists at Cleveland Clinic provided annotations for the four physiologically meaningful categories: \textit{None} (normal filling without involuntary contractions), \textit{Overactivity} (spontaneous detrusor contractions indicating overactive bladder syndrome), \textit{Voiding} (intentional or reflexive bladder emptying with sustained pressure rise), and \textit{Abdominal Activity} (non-detrusor pressure fluctuations from patient movement or external strain).

\section{Experiments and Results}

For evaluation, we compare two of our model variants: a \textit{segment-level transformer model} that performs transformer processing only within each 8-sample segment with no inter-segment communication, and a \textit{streaming transformer model} that additionally incorporates cross-segment caching with representations from the previous 8 segments. 
Both models are optimized using the AdamW optimizer~\cite{loshchilov2017decoupled}, trained for 100 epochs. The learning rate is set to $10^{-6}$ for the streaming transformer model and $10^{-3}$ for the segment-level transformer.
Note that the baseline results under the single-label setting are directly reported from prior work~\cite{majerus2024realtime} and are included here solely for comparative purposes. As the baseline model does not support multi-label classification, all multi-label results presented are a contribution of our framework. Additionally, all other results, including class-wise single-label metrics for both the segment-level and streaming transformer models, are obtained from our implementation using a unified evaluation pipeline to ensure consistency across experimental settings.

For the single-label classification task, we convert the original multi-label annotations by treating each bladder activity state as a binary indicator at the sample level and apply a threshold-based voting scheme within each segment to assign the dominant class label.
As shown in Table~\ref{table:performance}, the streaming transformer achieves an overall accuracy of 82.12 \%, outperforming both the baseline ANN (78.8 \%) and the segment-level transformer (79.2 \%).
On the VOID class, the streaming transformer also attains the highest accuracy (89.0 \%) and F1 score (67.5 \%), compared with 87.2 \% / 63.5\% for the segment-level transformer and 85.2 \% / 62.4\% for the baseline.
The streaming model therefore delivers a better precision–recall balance.
In the multi-label setting, the advantage of the streaming transformer becomes even more pronounced: it reaches an overall accuracy of 86.9 \%, versus 82.56 \% for the baseline and 84.34 \% for the segment-level transformer.
Consistent gains are observed across all classes.
On the DO class, the baseline and segment-level transformer still struggle (F1 = 6.1 \% and 12.1 \%, respectively), while the streaming transformer raises F1 to 12.5 \% by recovering more true positives.
Similarly, for VOID, the streaming model improves accuracy from 87.1 \% (baseline) and 91.0 \% (segment-level) to 92.1 \%, again reflecting a superior precision–recall trade-off.
We also performed ablation studies by varying segment length and model dimensionality. Reducing the embedding dimension below 32 harmed accuracy, while increasing beyond 64 yielded diminishing returns. Segment sizes larger than 8 yielded marginal performance gains but introduced additional latency, suggesting that 8-sample segments offer a favorable trade-off. 

\section{Conclusion}

We presented a single-layer streaming transformer architecture for real-time bladder pressure classification, incorporating segment-wise attention and cross-segment memory caching to balance temporal modeling with computational efficiency. Our approach processes wavelet-transformed signals in fixed 8-sample segments while maintaining lightweight context through pooled embeddings, enabling native multi-label classification of overlapping physiological states. The model achieves 83.25\% classification accuracy compared to 78.8\% for previous ANN-based methods, while requiring less than 1~ms inference time per segment on a NVIDIA Jetson edge GPU. With a complete footprint under 50~kB, the architecture is deployable on ARM Cortex-M7 microcontrollers with real-time 10~Hz throughput, making it suitable for next-generation wearable or implantable urological monitoring systems. The fixed computational graph and elimination of recurrent dependencies enable efficient hardware acceleration on edge devices. 
Future directions include adaptive segmentation strategies and extension of our transformer architecture to multimodal physiological inputs to support richer context-aware diagnostics in autonomous bladder management systems.


\balance

\bibliographystyle{IEEEtran}
\bibliography{references}

\end{document}